# Structural and magnetic properties of $Mn_{3-x}Cd_xTeO_6$ (x = 0, 1, 1.5 and 2)

S. A. Ivanov[ab], R. Mathieu[b*], P. Nordblad[b], E. Politova[a], R. Tellgren[c], C. Ritter[d], V. Proidakova[a]

a- Department of Inorganic Materials, Karpov' Institute of Physical Chemistry, Vorontsovo pole, 10 105064, Moscow K-64, Russia
b- Department of Engineering Sciences, Uppsala University, Box 534, SE-751 21 Uppsala, Sweden
c- Department of Materials Chemistry, Uppsala University, Box 538, SE-751 21 Uppsala, Sweden
d- Institute Laue Langevin, Grenoble, France

## Abstract

$Mn_3TeO_6$ exhibits a corundum-related $A_3TeO_6$ structure and a complex magnetic structure involving two magnetic orbits for the Mn atoms [*]. $Mn_{3-x}Cd_xTeO_6$ (x=0, 1, 1.5 and 2) ceramics were synthesized by solid state reaction and investigated using X-ray powder diffraction, electron microscopy, calorimetric and magnetic measurements. $Cd^{2+}$ replaces $Mn^{2+}$ cations without greatly affecting the structure of the compound. The Mn and Cd cations were found to be randomly distributed over the $A$-site. Magnetization measurements indicated that the samples order antiferromagnetically at low temperature with a transition temperature that decreases with increasing Cd doping. The nuclear and magnetic structure of one specially prepared $^{114}Cd$ containing sample: $Mn_{1.5}{}^{114}Cd_{1.5}TeO_6$, was studied using neutron powder diffraction over the temperature range 2 to 295 K. $Mn_{1.5}{}^{114}Cd_{1.5}TeO_6$ was found to order in an incommensurate helical magnetic structure, very similar to that of $Mn_3TeO_6$ [*]. However, with a lower transition temperature and the extension of the ordered structure confined to order 240(10) Å.
[*] S. A. Ivanov et al. Mater. Res. Bull. 46 (2011) 1870.

## 1. Introduction

A class of potential single-phase multiferroic materials possesses geometrically frustrated spin networks, which prevent the formation of conventional collinear spin structures. Instead, an incommensurate magnetic structure with spiral spin order may be established, and the magnetic phase transition to this non-centrosymmetrically ordered magnetic state can induce a spontaneous electric polarization via the spin-orbit coupling [1]. $Mn_3TeO_6$ crystallizes in a pseudocubic, rhombohedral structure with fairly closed packed oxygen anions [2], and orders magnetically at low temperatures in an incommensurate helical structure [3].
Substitution of $Cd^{2+}$ for $Mn^{2+}$ in $Mn_3TeO_6$ alters the electronic structure significantly but not the crystal structure. $Mn_{3-x}Cd_xTeO_6$ (MCTO) solid solutions are known as antiferroelectrics [4, 5]. Compounds with x=1 and 2 were first prepared as powder by *Bayer* [6, 7] for studies that mainly were focused on crystallographic aspects. Later, *Kosse* [4, 5] reported on single-phase ceramics of compositions x = 0, 1 and 2, which were described as antiferroelectrics with $T_{AFE}$ = 510 K, 570 K and 480 K respectively. The understanding of these dielectric anomalies is a matter of debate and the magnetic properties of the MCTO system have not yet been clarified.

* Corresponding author:
*E-mail address*: roland.mathieu@angstrom.uu.se

In this article we report results from X-ray diffraction and magnetic and heat capacity measurements of MCTO which indicate the development of long-range magnetic order at low temperatures. In addition, results from the first neutron diffraction experiments on MCTO ($Mn_{1.5}{}^{114}Cd_{1.5}TeO_6$) are reported. This compound is found to order magnetically below 6 K in an incommensurate magnetic structure with different types of Mn-chains. The propagation vector (k = [0, 0, 0.435(2)]) is modulating the spins to a helical spiral with a length of 24.9 Å and a turn angle of 156.6°. Interestingly, this study contains one of the few reports of refinement of a magnetic structure from NPD data of a compound containing Cd ions.

## 2. Experimental

### 2.1 Sample preparation
High quality ceramic samples of MCTO were prepared by a conventional solid-state ceramic route. High purity $Mn_2O_3$, carbonate $CdCO_3$, $TeO_3$ and telluric acid $H_2TeO_4 \cdot 2H_2O$ were used as starting materials. The raw materials were weighed in appropriate proportions for the different $Mn_{3-x}Cd_xTeO_6$ compositions. The homogenized stoichiometric mixtures were calcined at 570 K for 7 h, ground into fine powders, pressed into tablets and annealed in closed Pt crucibles again several times with a temperature interval of 100 K up to 1270 K with intermediate milling. The maximal synthesis temperatures were below the melting points of the solid solutions. The compounds were quenched by removing the samples from the furnace immediately after sintering. Single phase samples with compositions $Mn_2CdTeO_6$ and $MnCd_2TeO_6$ could be prepared by hot pressing at T=1170 K and P=2.67 MPa. An isotope-enriched ($^{114}Cd$, 99% richness, Trace Sciences International) CdO was used as the cadmium source for preparation of $Mn_{1.5}{}^{114}Cd_{1.5}TeO_6$.

### 2.2. Sample characterisation
The chemical compositions of the prepared ceramic samples were determined by energy-dispersive spectroscopy (EDS) using a JEOL 840A scanning electron microscope and INCA 4.07 (Oxford Instruments) software.

The phase identification and purity of the powder samples were checked from X-ray powder diffraction patterns (XRPD) obtained with a Bruker D-5000 diffractometer using Cu $K_\alpha$ radiation. For that purpose, the ceramic MCTO samples were crushed into powder in an agate mortar and suspended in ethanol. A Si substrate was covered with several drops of the resulting suspension, leaving randomly oriented crystallites after drying. The XRPD data for Rietveld analysis were collected at room temperature on a Bruker D8 Advance diffractometer (Vantec position-sensitive detector, Ge-monochromatized Cu $K_\alpha$ radiation, Bragg-Brentano geometry, DIFFRACT plus software) in the $2\theta$ range 10-152° with a step size of 0.02° and a counting time of 15 s per step. The slit system was selected to ensure that the X-ray beam was completely within the sample for all $2\theta$ angles.

Polycrystalline MCTO materials were characterized by second harmonic generation (SHG) SHG measurements in reflection geometry, using a pulsed Nd:YAG laser ($\lambda$=1.064 μm). The SHG signal $I_{2\omega}$ from the sample was measured relative to an α-quartz standard at room temperature in the Q-switching mode with a repetition rate of 4 Hz.

The magnetization experiments were performed on ceramic samples in a Quantum Design MPMSXL 5 T SQUID magnetometer. The magnetization (M) was recorded as a function of temperature T in the interval 2-300 K in 50 Oe and 5000 Oe magnetic fields using zero-field-cooled (ZFC) and field-cooled (FC) protocols.

Specific heat measurements between 2 K and 50 K were performed using a relaxation method on a Physical Properties Measurement System (PPMS6000) from Quantum Design Inc**.**

## 2.3. Neutron powder diffraction

Because the neutron scattering lengths of Mn, Cd and Te are very different, the chemical composition can be observed by neutron powder diffraction (NPD) with good precision. The neutron scattering length of oxygen is comparable to those of the cations and NPD provide accurate information on its position and stoichiometry. However, natural cadmium (containing 12% $^{113}$Cd) is an extremely strong absorber of neutrons ($\sigma_{abs}$ = 2520 b/atom). This makes neutron studies of materials containing even moderate levels of cadmium impossible. Nevertheless, some isotopes of cadmium have negligible neutron absorption characteristics, e.g. $^{114}$Cd ($\sigma_{abs}$ = 0.34 b/atom), making materials substituted with this isotope alone suitable for neutron diffraction research. We used isotopic enriched cadmium $^{114}$Cd (99.2%) to fabricate one composition of MCTO: $Mn_{1.5}Cd_{1.5}TeO_6$. The neutron diffraction experiments on this $Mn_{1.5}{}^{114}Cd_{1.5}TeO_6$ sample were performed at the Institute Laue-Langevin (Grenoble, France) on the powder diffractometer D1A ($\lambda$ = 1.9095(1) Å) in the $2\theta$-range 10–156.9° with a step size of 0.1°. The powdered sample was inserted in a cylindrical vanadium container. A helium cryostat was used to collect data in the temperature range 1.5-295 K. Nuclear and magnetic refinements were performed by the Rietveld method using the *FULLPROF* software [8]. The scattering length for the $^{114}$Cd isotope was taken as 7.5 fm [9]. The diffraction peaks were described by a pseudo-Voigt profile function, with a Lorentzian contribution to the Gaussian peak shape. A peak asymmetry correction was made for angles below 35°($2\theta$). Background intensities were estimated by interpolating between up to 40 selected points (low temperature NPD experimental data) or described by a polynomial with six coefficients. Analysis of the coordination polyhedra of the cations was performed using the *IVTON* software [10].

The magnetic propagation vector was determined from the peak positions of the magnetic diffraction lines using the K-search software which is included in the *FULLPROF Suite* package [8].

Magnetic symmetry analysis was then made using the program *BASIREPS* [11] which calculates the allowed irreducible representations and their basis vectors. The proposed models were one by one tested against the measured data. The variant for which the structural refinement was stable and the reliability factors at a minimum was chosen as the final model. A special software for profile fitting (program *PROFIT* [12]) was used for analysis of the broadening of magnetic peaks.

## 3. Results

### 3.1 Structural Characterization

According to the X-ray diffraction data, the MCTO sample formation starts in temperature interval of 850-920 K, and after annealing at 970-1050 K single phase samples were prepared. Elemental analyses, performed on 20 different crystallites, showed that the metal compositions of the MCTO samples are close to the expected ratios and permit to conclude that the sample compositions are the nominal ones (within the instrumental resolution (0.05)). The oxygen contents, as determined by thermogravimetric analysis, are also in agreement with the proposed anion composition.

The microstructure of the obtained powders, observed by scanning electron microscopy, reveals uniform grain distribution. Second harmonic generation (SHG) measurements at room temperature gave negative results, thus testifying that at this temperature the MCTO compounds probably possess a centrosymmetric crystal structure.

The first crystallographic characterization of MCTO compounds was performed by XRPD analysis at room temperature which showed that the prepared samples are single phase and show resolution limited diffraction peaks without any splitting or extra reflections. The room temperature XRPD patterns for all prepared compounds could be indexed on the basis of hexagonal unit cell parameters (see Table 1). These values are in a reasonable agreement with the lattice parameters obtained earlier by *Kosse* in [4, 5] and *Bayer* in [6, 7]. The XRD powder patterns and lattice metrics of the $Mn_{1.5}Cd_{1.5}TeO_6$ samples with natural Cd ions and $^{114}Cd$ isotope enriched ions were similar.

XRPD patterns could be successfully refined by the Rietveld method using s.g. $R\bar{3}$ (see Table 1). The possible cationic ordering of $Cd^{2+}$ over the sites of $Mn^{2+}$ is charge and size dependent and cations of the same or closely equal size tend to distribute statistically (in a disordered manner) on the same site. Indeed, it was found that $Mn^{2+}$ (r=0.83 Å) and $Cd^{2+}$ (r=0.95 Å) prefer to occupy the same octahedral 18f position. The Mn/Cd-O and Te-O bond lengths calculated from the refined lattice parameters and atomic coordinates are in good agreement with earlier observed for other Mn-based tellurates [13].

Furthermore, the corresponding bond valence sum calculations are consistent with the presence of $Mn^{2+}$, $Cd^{2+}$, $Te^{6+}$ and $O^{2-}$ ions. The steady increase of the lattice parameters and unit cell volume within the solid solution domain was ascribed to the substitution of $Mn^{2+}$ cations by larger sized $Cd^{2+}$ ones as the composition progresses between x=0 and x=2. The final results from the Rietveld refinements with observed, calculated and difference plots for XRPD pattern of $Mn_{1.5}{}^{114}Cd_{1.5}TeO_6$ at 295 K are shown in Figure 1.

### 3.2 Magnetic properties

Figure 2 shows the low-temperature magnetic susceptibility of $Mn_{1.5}{}^{114}Cd_{1.5}TeO_6$. Two anomalies, near 10K and 6K, can be observed in the ZFC-curve recorded in low magnetic field (panel (a)). The higher temperature anomaly is not observed in the curves recorded in a larger magnetic field (panel (b)), suggesting that it originates from a weak spontaneous magnetic moment, originating from a faint impurity phase not detectable in our XRD and NPD data but readily observed in the magnetization data. The lower temperature anomaly is robust to magnetic fields and indicates an antiferromagnetic type of ordering below 6K. Also plotted in panel (a) are ZFC/FC susceptibility curves for $Mn_{3-x}Cd_xTeO_6$ samples with x=0 and x=1. Antiferromagnetic transitions can be distinguished from weak anomalies at about 23 K for the x=0 compound and near 11 K for x=1. As seen in the inset of panel (b), the susceptibility data of $Mn_{1.5}{}^{114}Cd_{1.5}TeO_6$ at high temperatures follows a Curie-Weiss law $M/H(T)=C/(T+\theta)$ with $\theta \sim -48$ K, indicating relatively strong antiferromagnetic interaction combined with frustration, considering the much lower magnitude of $T_N$ compared to $\theta$ [14]. The derived effective Bohr magneton number $p$ per Mn ion amounts to 5.89 $\mu_B$, which is close to the expected value for $Mn^{2+}$ in a $3d^5$ configuration S=5/2, L=0 ($p$=5.92 $\mu_B$).

The temperature dependence of the heat capacity C (plotted as C(T)/T in the main frame and C(T) in the inset) of $Mn_{1.5}{}^{114}Cd_{1.5}TeO_6$ is shown in Figure 3. The increase of the heat capacity below 30K suggests a magnetic contribution below that temperature. A peak is observed at lower temperatures, with maximum near T=6K, i.e. in the vicinity of the anomaly recorded in the magnetization measurements. The magnetization and heat capacity data suggest antiferromagnetic ordering below 6K in $Mn_{1.5}{}^{114}Cd_{1.5}TeO_6$. For comparison we have included heat capacity on $Mn_{1.5}Cd_{1.5}TeO_6$ in Figure 3, which behaves quite similarly to $Mn_{1.5}{}^{114}Cd_{1.5}TeO_6$. The difference in ordering temperature between the two samples of nominally identical composition can be explained by a strong sensitivity of $T_N$ on the amount of frustration in the sample, which in turn is very sensitive to any deviation from random distribution of $Mn^{2+}$ and $Cd^{2+}$ in the samples.

### 3.3 Neutron powder diffraction at room temperature

Rietveld refinements of the NPD data at 295 K confirmed the XRD result that the crystal structure of the $Mn_{1.5}{}^{114}Cd_{1.5}TeO_6$ sample is the same as that of $Mn_3TeO_6$ [2, 3]. No vacancies were observed in the cationic or in the anionic substructures. Accordingly, the Mn/Cd oxidation state can be assumed to be +2. The atomic coordinates and other relevant parameters are gathered in Table 2. Selected bond lengths and angles are listed in Table 3. The crystal structure of $Mn_{1.5}{}^{114}Cd_{1.5}TeO_6$ is isotypical to $Mn_3TeO_6$ and can be derived from a close packing of strongly distorted hexagonal oxygen layers parallel to (001), with Mn/Cd and two distinct Te atoms in the octahedral interstices (see Figure 4). Both $TeO_6$ octahedra exhibit 3-fold symmetry and are fairly regular, with a Te-O distance of 1.92-1.94 Å (Table 3), which is in good agreement with the average Te-O distances of other oxotellurates [13]. Each $TeO_6$ octahedron shares edges with six $Mn/CdO_6$ octahedra but none with other $TeO_6$ octahedra. Each $Mn/CdO_6$ octahedron shares four edges with adjacent $Mn/CdO_6$ octahedra, one edge with a $Te(1)O_6$ and another edge with the $Te(2)O_6$ octahedron. The shared edges of the $TeO_6$ octahedra have somewhat shorter O-O distances than the non-shared edges. This is maybe due to the high valence of tellurium which has the tendency to keep as far away as possible from the $Mn^{2+}/Cd^{2+}$ cations. Each of the two crystallographically independent O atoms is coordinated by one Te and three Mn/Cd cations in a distorted tetrahedral manner. The $Mn/CdO_6$ octahedron is considerably distorted which is reflected by the variation of the Mn/Cd-O distances between 2.092(6) to 2.493(6) Å (Table 3).

**3.4 Low Temperature Neutron Powder Diffraction**
Low temperature NPD measurements on the $Mn_{1.5}{}^{114}Cd_{1.5}TeO_6$ sample were performed at 30 K and at 2 K. The diffraction profiles are shown in Figure 5. At 30 K, the NPD pattern shows the peaks expected from the room-temperature crystal structure and, in addition, some diffuse scattering over a notable 2θ range centered around 20°. This diffuse scattering may indicate the presence of short-range antiferromagnetic correlations developing above $T_N$.
The 2 K NPD pattern shows the presence of additional small broad peaks of magnetic origin (see Figures 5 and 6) that has to be taken into account in the refinement. A magnetic structure for $Mn_{1.5}{}^{114}Cd_{1.5}TeO_6$ was refined taking the magnetic structure reported earlier for $Mn_3TeO_6$ [3] as a starting point and, in fact, this same model adequately describes also the magnetic structure of the magnetically diluted sample, $Mn_{1.5}{}^{114}Cd_{1.5}TeO_6$. The magnetic Bragg peaks could not be indexed using simple multiples of the crystallographic unit cell, implying that the magnetic structure is incommensurate with respect to the nuclear structure of the crystal. The value of the propagation vector, which was able to index all the magnetic satellite peaks was $k = [0, 0, 0.435(2)]$. As in the case of $Mn_3TeO_6$ [3], the program "BasIreps" [11] was used to find allowed symmetry couplings in the form of irreducible representations and their respective basis vectors. The refinement required a representation with both real and imaginary components which implies a helical type of arrangement of the spins [15]. Table 4 shows the refined coefficients for the resulting magnetic structure involving two Mn orbits; Mn(1) and Mn(2). Due to the hexagonal axes with an angle of 120° the resulting spin structure has an elliptical envelope perpendicular to the propagation vector. The structure is illustrated in Figure 7: the helical spin evolution of Mn(1) along the c-axis can be seen in Figure 7a. The first orbit Mn(1) has no spin component in the c-direction, whereas the second orbit Mn(2) has its spin component only in the c-direction with no components within the basal plane, as shown in Figure 7b. The sine wave character of the Mn(2) moments is depicted in Figure 7c.
The magnetic moment values are in the range 1.6(2) $\mu_B$ – 2.9(2) $\mu_B$ for the Mn(1) spins which sees the moments exclusively in the hexagonal basal plane and between 0.5(2) $\mu_B$ and 2.1(2) $\mu_B$ for the Mn(2) spins belonging to the second orbit with sinusoidal component along the c-axis. These moment values agree well with the fact that the A-site is occupied by equal amounts of $Mn^{2+}$ and $Cd^{2+}$ ions and that the corresponding moment values for the undoped, $Mn_3TeO_6$, sample [3] are about a factor of 2 larger.
The value of the incommensurate magnetic propagation vector $k = [0, 0, 0.435(2)]$ leads to a spiral with a length of 24.9 Å and a turn angle of 156.6° between spins in neighbouring cells in the c-direction. Along the a and b axes no incommensurate modulation is present and the repeat unit is 8.99 Å.

Position and profile parameters of several magnetic reflections were obtained by peak profile fitting. The width of the peaks has been used for an estimation of the magnetic correlation length at 2 K. If we picture the ordered magnetic structure in $Mn_{1.5}{}^{114}Cd_{1.5}TeO_6$ as a simple set of ordered domains, the domain size can be estimated using the *Scherrer* formula [16] based on the reciprocal relation between domain size $D$(Å) and peak half-width broadening *FWHM* (degrees) according to

$$D = \lambda \times 57.3 / cos\theta \times FWHM$$

By taking the actual experimental parameters (after correction for instrumental resolution), a characteristic value of 240(10) Å for magnetic domains was calculated. This is consistent with the heat capacity data: The peak in the heat capacity near 6K is relatively broad, suggesting that the antiferromagnetic correlation length, albeit large, might remain finite. In comparison, the size of the crystallites forming the samples is from XRD and NPD data estimated to be of order 0.1 μm.

## 4. Discussion

The antiferromagnetic superexchange along a pathway Mn-O-Mn may be expected to give rise to the strongest magnetic interaction in the $Mn_3TeO_6$ type structure, causing a G-type antiferromagnetic structure [17] (where the Mn cations are AFM coupled with the six neighboring cations) at low temperature. However, it is important to note that in this structure there are 3 Mn/Cd atoms within a ring (at the same height in the c direction) having equivalent individual bonds giving rise the classical case of frustration in a triangle (see Figure 8), and preventing a simple antiferromagnetic order to occur. Instead we observe the described helical order at a transition temperature much lower than the strength of the antiferromagnetic interaction indicates. It is also striking that the same helical structure describes the magnetic structure of the diluted compounds all the way to the x=1.5 substitution, $Mn_{1.5}Cd_{1.5}TeO_6$. The octahedral *A*-site in $Mn_{1-x}Cd_xTeO_6$ is occupied to *1, 2/3, 1/2* and *1/3* (x=0, 1, 1.5 and 2, respectively) with the magnetic ($Mn^{2+}$) cation. The magnetic transition temperature correspondingly decreases with magnetic dilution from 23 K for x=0 to 11 K for x=1 and 6 K for x=1.5. This decay directly relates to the dilution effect and a percolation limit [18] that from the confined range of the magnetic order for the x=1.5 sample appears close to 0.5 (the percolation limit for a triangular lattice).

Most frustrated compounds which undergo true phase transitions to a long range ordered state adopt non-collinear or so-called compromise spin configurations [19]. The canonical example is the triangular lattice in which the three spins interact antiferromagnetically [20] and there is more than one way to achieve an AFM ground state [21]. A corresponding triangular configuration can, as mentioned above, be found for the Mn (Mn/Cd) ions in $Mn_{3-x}Cd_xTeO_6$. The principal geometric parameters for qualitative understanding of magnetic interactions are the Mn-Mn distances and the Mn-O-Mn angles. For superexchange interaction the most favorable metal-oxygen-metal angle is 180° and a near 90° angle is most unfavorable. According to the *Goodenough - Kanamori* rules [22, 23], a decreasing Mn-O-Mn angle results in a monotonic decrease from a strong antiferromagnetic coupling at an 180° bond to a weak ferromagnetic coupling at a 90° bond. In the studied structural family, Mn cations order in alternating layers along the trigonal axis which are rather isolated from one another, and no 180° connectivity (as in the case of perovskite structure) is to be expected. Indeed, in the case of a corundum-related structure the bonding angles of Mn-O-Mn are between 94° and 117°, which causes geometrical frustration and weakened super-exchange interaction [24].

The relative variation of lattice parameters with substitution is related to the ionic radius of the substituting cation: both *a* and *c* shows a linear increase with doping by $Cd^{2+}$ ions, which have a larger radius than $Mn^{2+}$. If one takes into account the values of the polyhedral distortions for all samples including $Mn_{1.5}{}^{114}Cd_{1.5}TeO_6$ at room temperature (see Table 5), possible ferroic properties in these compounds may be connected with the strongly distorted A- sublattice, where Mn/Cd cations are significant displaced from the polyhedral centers. These cation shifts are increased with Cd-doping, but it is difficult to find a simple explanation for the different values of $T_{AFE}$ [4, 5] based on only crystallochemical considerations at temperatures relatively far from the phase transition. The development of coordination polyhedra with Cd-doping suggests an increase of volume and compressibility for the *A*-type octahedra. It is worth to notice that the volumes of the Te(1) and Te(2) polyhedra are smaller and they are negligibly distorted in comparison with the $(Mn/Cd)O_6$ octahedra.

## 4. Conclusions

All studied $Mn_{3-x}Cd_xTeO_6$ samples adopt a corundum-related trigonal structure at room temperature, and we have shown that $Mn_{1.5}{}^{114}Cd_{1.5}TeO_6$ retains this symmetry down to 2 K. Some key features of the structure are a close packing of strongly distorted hexagonal oxygen layers parallel to (001), with Mn/Cd and two distinct Te atoms in the octahedral interstices. Both $TeO_6$ octahedra are fairly regular but the $Mn/CdO_6$ octahedron is considerably distorted. The relatively long distances between $Mn^{2+}$ ions and frustration effects are possibly responsible for the low ordering temperature of the parent compound $Mn_3TeO_6$; the rapid decrease of the ordering temperature with Cd-doping is caused by further increased distance between the $Mn^{2+}$ ions and, of course, magnetic dilution. Below 6 K, $Mn_{1.5}{}^{114}Cd_{1.5}TeO_6$ enters an incommensurate magnetically ordered state in which the spins form helical spirals of length 24.9 Å and a turn angle of 156.6°. The spin-spin correlation length derived from peak shape fitting of neutron diffraction data at 2 K was estimated to 240(10) Å; a confinement that may be assigned to a Mn occupancy of 0.5 on the A-site, which would be at the percolation limit assuming a purely triangular interaction pattern.

## Acknowledgements

Financial support from the Swedish Research Council (VR), the Göran Gustafsson Foundation and the Russian Foundation for Basic Research is gratefully acknowledged. We also gratefully acknowledge the support from N. Sadovskaya and S. Stefanovich during the EDS cation analysis and second harmonic generation testing.

Table 1 Summary of the results of the structural refinements of the Mn$_{3-x}$Cd$_x$TeO$_6$ samples using XRPD data at 295 K.

| $x$ | 0 | 1 | 1.5 | 2 |
|---|---|---|---|---|
| a[Å] | 8.8675(1) | 8.9583(1) | 9.0024(1) | 9.0814(1) |
| c[Å] | 10.6731(2) | 10.7999(2) | 10.8593(2) | 10.9749(2) |
| s.g. | $R$-3 | $R$-3 | $R$-3 | $R$-3 |
| c/a | 1.204 | 1.205 | 1.206 | 1.208 |
| V[Å$^3$] | 726.7 | 750.5 | 762.1 | 783.8 |
| T$_{AFE}$,K | 510 | 570 | 530 | 480 |
| **Mn/Cd** | | | | |
| $x$ | 0.0384(5) | 0.0381(4) | 0.0380(6) | 0.0347(5) |
| $y$ | 0.2643(5) | 0.2629(5) | 0.2649(5) | 0.2617(4) |
| $z$ | 0.2131(4) | 0.2127(3) | 0.2128(4) | 0.2145(4) |
| $B$[Å]$^2$ | 0.74(1) | 0.77(6) | 0.83(6) | 0.81(6) |
| **Te(1)** | | | | |
| x | 0 | 0 | 0 | 0 |
| y | 0 | 0 | 0 | 0 |
| z | 0.5 | 0.5 | 0.5 | 0.5 |
| $B$[Å]$^2$ | 0.38(3) | 0.44(4) | 0.56(4) | 0.42(3) |
| **Te(2)** | | | | |
| $x$ | 0 | 0 | 0 | 0 |
| $y$ | 0 | 0 | 0 | 0 |
| $z$ | 0.5 | 0 | 0 | 0 |
| $B$[Å]$^2$ | 0.31(3) | 0.30(4) | 0.33(3) | 0.35(4) |
| **O(1)** | | | | |
| $x$ | 0.0308(6) | 0.0369(4) | 0.0300(6) | 0.0307(5) |
| $y$ | 0.1964(4) | 0.1996(5) | 0.1961(5) | 0.1963(4) |
| $z$ | 0.4029(3) | 0.3948(4) | 0.4030(3) | 0.4029(5) |
| $B$[Å]$^2$ | 0.70(5) | 0.86(6) | 0.94(5) | 0.84(6) |
| **O(2)** | | | | |
| $x$ | 0.1825(4) | 0.1820(4) | 0.1833(5) | 0.1829(4) |
| $y$ | 0.1560(6) | 0.1522(5) | 0.1561(5) | 0.1562(6) |
| $z$ | 0.1104(3) | 0.1134(4) | 0.1113(3) | 0.1106(5) |
| $B$[Å]$^2$ | 0.82(5) | 0.94(6) | 0.83(5) | 0.97(6) |
| R$_p$,% | 4.74 | 4.65 | 5.16 | 5.34 |
| R$_{wp}$,% | 5.87 | 6.18 | 6.21 | 6.61 |
| R$_B$(%) | 4.28 | 3.97 | 4.61 | 4.15 |
| $\chi^2$ | 2.12 | 2.23 | 1.95 | 1.86 |

Table 2. Summary of the structural refinement results of the Mn$_{1.5}$$^{114}$Cd$_{1.5}$TeO$_6$ sample using NPD data.

| T,K | 2 | 30 | 295 |
|---|---|---|---|
| a[Å] | 8.9877(1) | 8.9882(1) | 9.0024(1) |
| c[Å] | 10.8397(2) | 10.8405(2) | 10.8593(2) |
| s.g. | *R*-3 | *R*-3 | *R*-3 |
| **Mn/Cd** | | | |
| *x* | 0.0416(5) | 0.0397(4) | 0.0366(6) |
| *y* | 0.2526(6) | 0.2551(5) | 0.2494(7) |
| *z* | 0.2205(4) | 0.2204(3) | 0.2296(4) |
| *B*[Å]$^2$ | 0.59(1) | 0.81(6) | 0.94(7) |
| **Te(1)** | | | |
| x | 0 | 0 | 0 |
| y | 0 | 0 | 0 |
| z | 0.5 | 0.5 | 0.5 |
| *B*[Å]$^2$ | 0.41(1) | 0.44(4) | 0.51(2) |
| **Te(2)** | | | |
| *x* | 0 | 0 | 0 |
| *y* | 0 | 0 | 0 |
| *z* | 0.5 | 0.5 | 0.5 |
| *B*[Å]$^2$ | 0.46(4) | 0.49(4) | 0.52(2) |
| **O(1)** | | | |
| *x* | 0.0286(4) | 0.0296(4) | 0.0293(6) |
| *y* | 0.1947(4) | 0.1946(3) | 0.1940(5) |
| *z* | 0.4041(3) | 0.4042(2) | 0.4041(3) |
| *B*[Å]$^2$ | 0.77(2) | 0.83(2) | 0.87(8) |
| **O(2)** | | | |
| *x* | 0.1821(5) | 0.1829(4) | 0.1823(5) |
| *y* | 0.1541(5) | 0.1547(4) | 0.1536(5) |
| *z* | 0.1075(3) | 0.1071(2) | 0.1066(3) |
| *B*[Å]$^2$ | 0.59(2) | 0.60(2) | 0.65(8) |
| R$_p$,% | 4.39 | 4.32 | 4.44 |
| R$_{wp}$,% | 5.82 | 5.65 | 6.07 |
| R$_B$(%) | 5.49 | 5.13 | 5.61 |
| R$_{mag}$,% | 11.4 | - | - |
| $\chi^2$ | 1.92 | 2.19 | 1.81 |



Table 3. Selected bond lengths [Å] from neutron powder data refinements of the $Mn_{1.5}{}^{114}Cd_{1.5}TeO_6$ sample at various temperatures.

| Bonds, Å | | 2 K | 30 K | 295 K |
|---|---|---|---|---|
| Mn/Cd | O1 | 2.046(6) | 2.053(6) | 2.092(6) |
| | O1 | 2.231(4) | 2.238(6) | 2.251(5) |
| | O1 | 2.531(4) | 2.523(6) | 2.493(6) |
| | O2 | 2.230(6) | 2.231(4) | 2.204(5) |
| | O2 | 2.266(4) | 2.272(5) | 2.261(6) |
| | O2 | 2.272(7) | 2.281(6) | 2.288(5) |
| Te(1) | O1 | 1.939(3) | 1.941(4) | 1.935(4) |
| Te(2) | O2 | 1.920(4) | 1.921(5) | 1.921(4) |

Table 4. Refined values of the coefficients $C_1$, $C_2$ and $C_3$ for the magnetic phase at 2 K ($C_1$ and $C_2$ are not orthogonal to each other but span $120°$). Mn(1) on x, y, z (0.041, 0.252, 0.221) belongs to orbit 1, Mn(2) on –x, -y, -z (0.959, 0.748, 0.779) belongs to orbit 2. Phase between the 2 orbits $\phi = 0.09(1)$.

| Atom | $C_1$ | $C_2$ | $C_3$ |
|---|---|---|---|
| Mn(1) | 2.38(9) | -2.38(9) | 0 |
| Mn(2) | 0 | 0 | 2.03(12) |

Table 5. Polyhedral analysis of $Mn_{3-x}Cd_xTeO_6$ at 295 K (cn - coordination number, x – shift from centroid, $\xi$- average bond distance with a standard deviation, V- polyhedral volume, $\omega$- polyhedral volume distortion.

| Cation | x | cn | x(Å) | $\xi$ (Å) | V(Å$^3$) | $\omega$ | Valence |
|---|---|---|---|---|---|---|---|
| Mn/Cd | 0 | 6 | 0.074 | 2.212+/-0.096 | 12.8(1) | 0.104 | 1.97 |
| | 1 | | 0.077 | 2.229+/-0.148 | 13.2(1) | 0.105 | 1.95 |
| | 1.5 | | 0.083 | 2.254+/-0.177 | 13.6(1) | 0.107 | 1.94 |
| | 2 | | 0.085 | 2.273+/-0.195 | 13.9(1) | 0.111 | 1.93 |
| Te(1) | 0 | 6 | 0 | 1.925+/-0.004 | 9.4(1) | 0.006 | 5.89 |
| | 1 | | 0 | 1.933+/-0.003 | 9.6(1) | 0.006 | 5.83 |
| | 1.5 | | 0 | 1.941+/-0.003 | 9.7(1) | 0.007 | 5.87 |
| | 2 | | 0 | 1.973+/-0.004 | 9.9(1) | 0.008 | 5.92 |
| Te(2) | 0 | 6 | 0 | 1.929+/-0.003 | 9.5(1) | 0.007 | 5.83 |
| | 1 | | 0 | 1.938+/-0.003 | 9.7(1) | 0.007 | 5.94 |
| | 1.5 | | 0 | 1.949+/-0.004 | 9.8(1) | 0.008 | 5.91 |
| | 2 | | 0 | 1.967+/-0.004 | 10.1(1) | 0.008 | 5.87 |



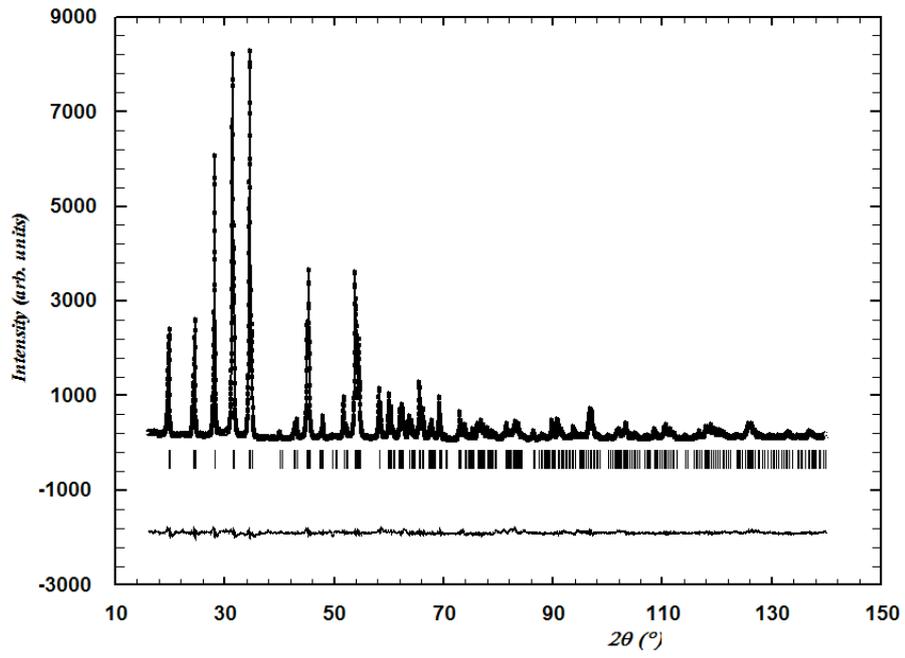

**Figure 1**



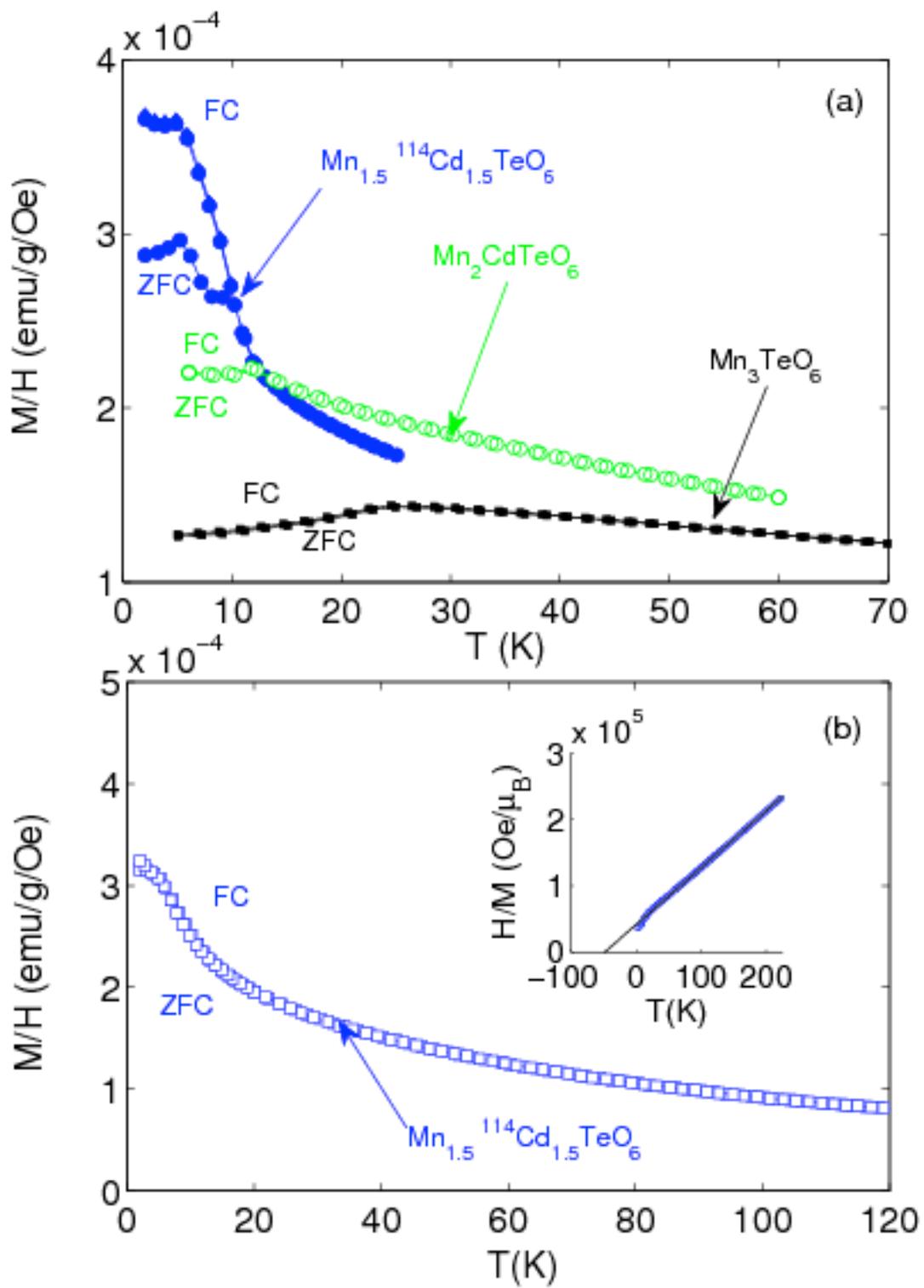

**Figure 2**



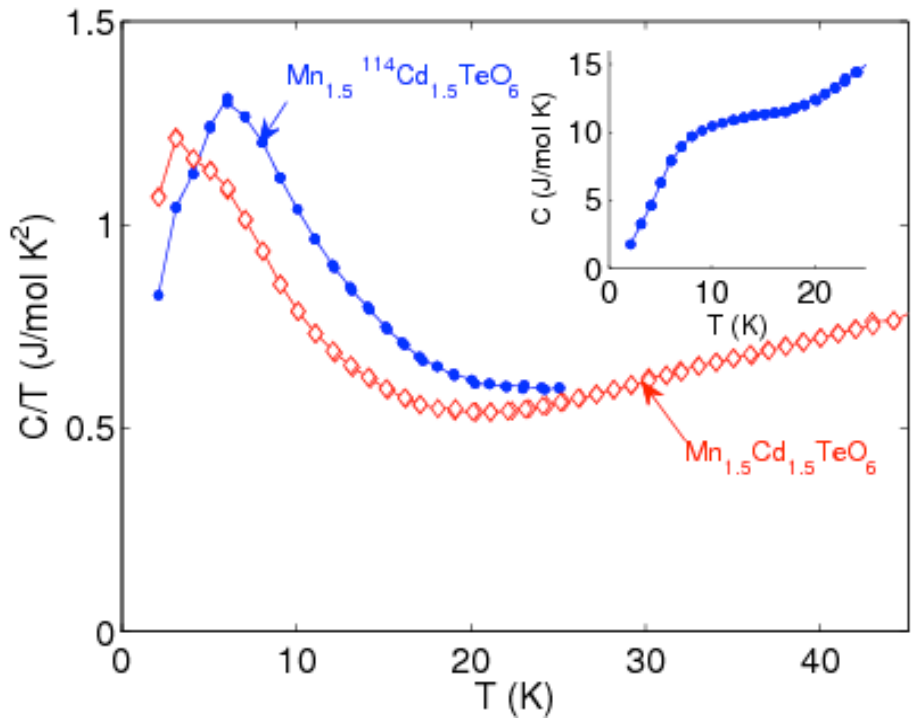

**Figure 3**



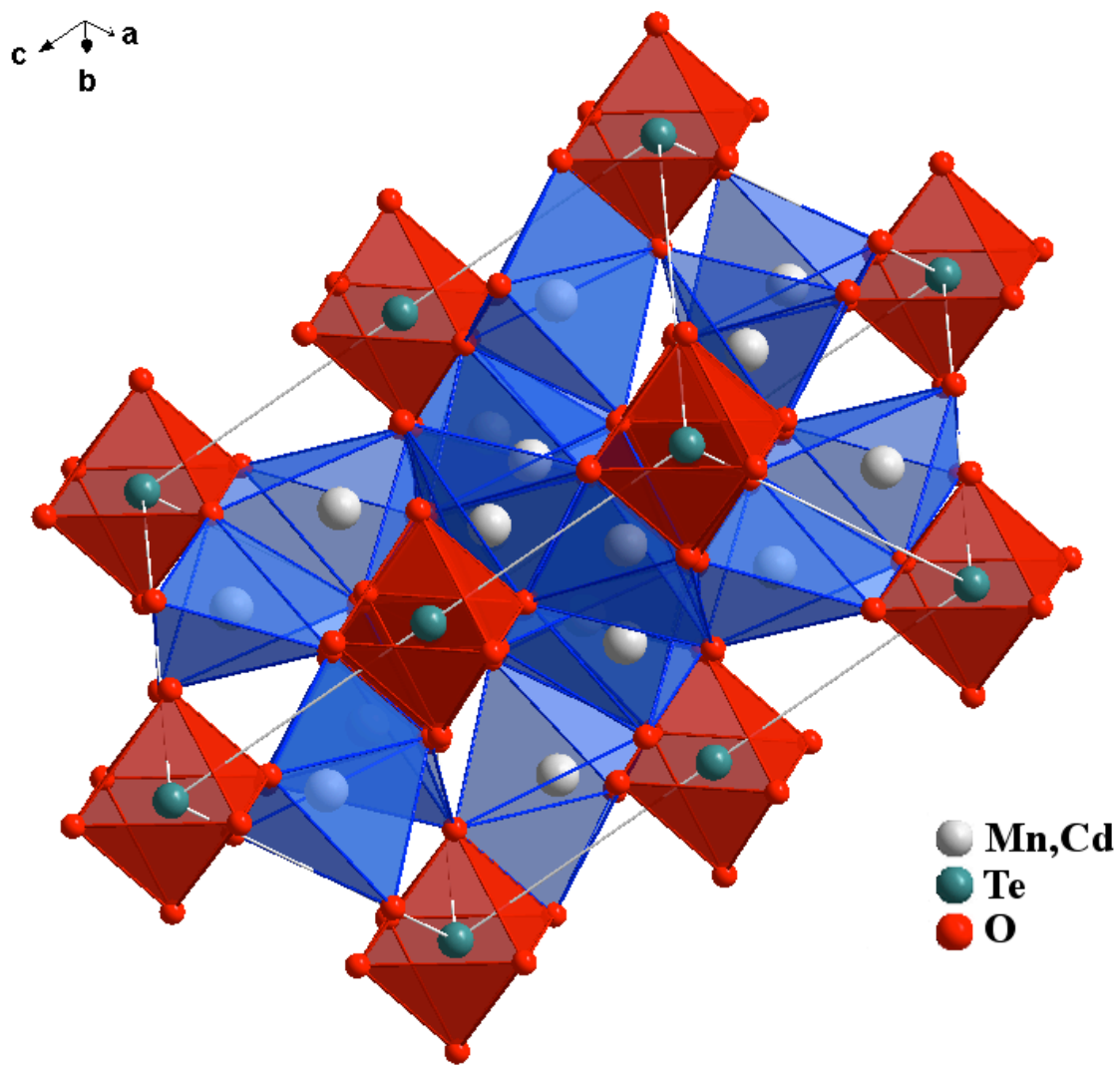

**Figure 4**



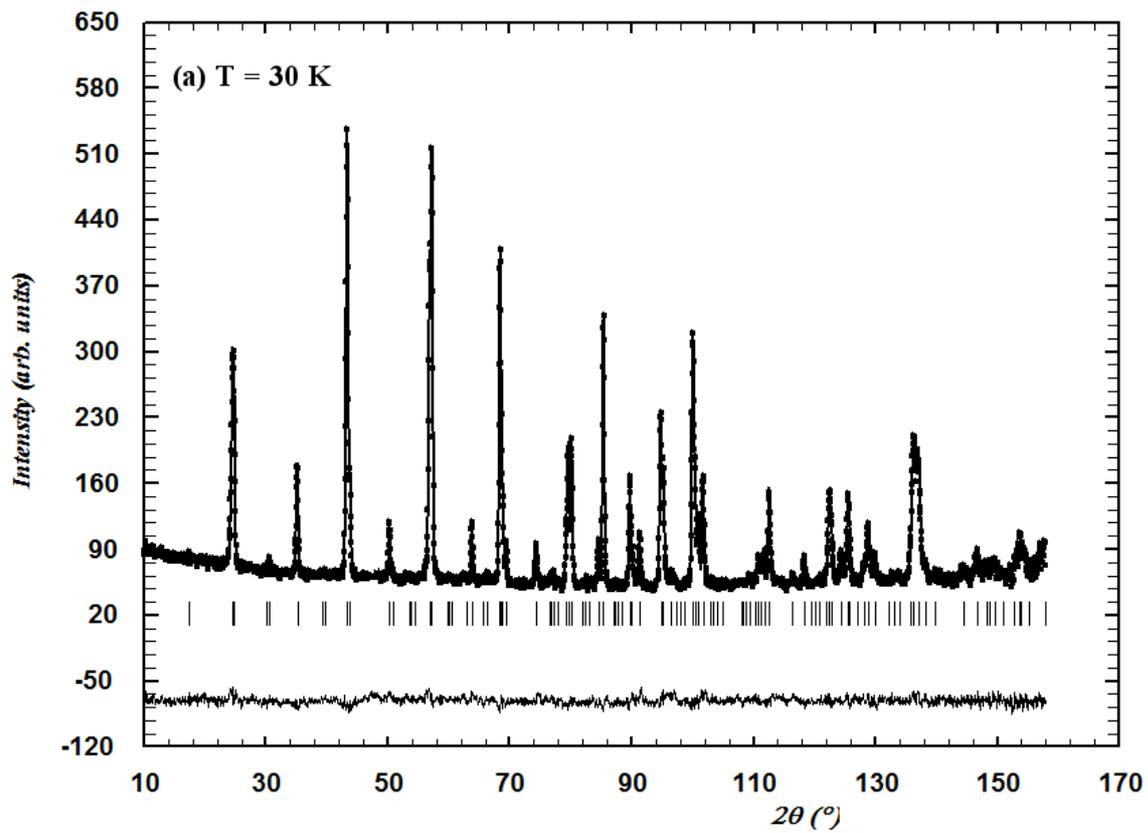

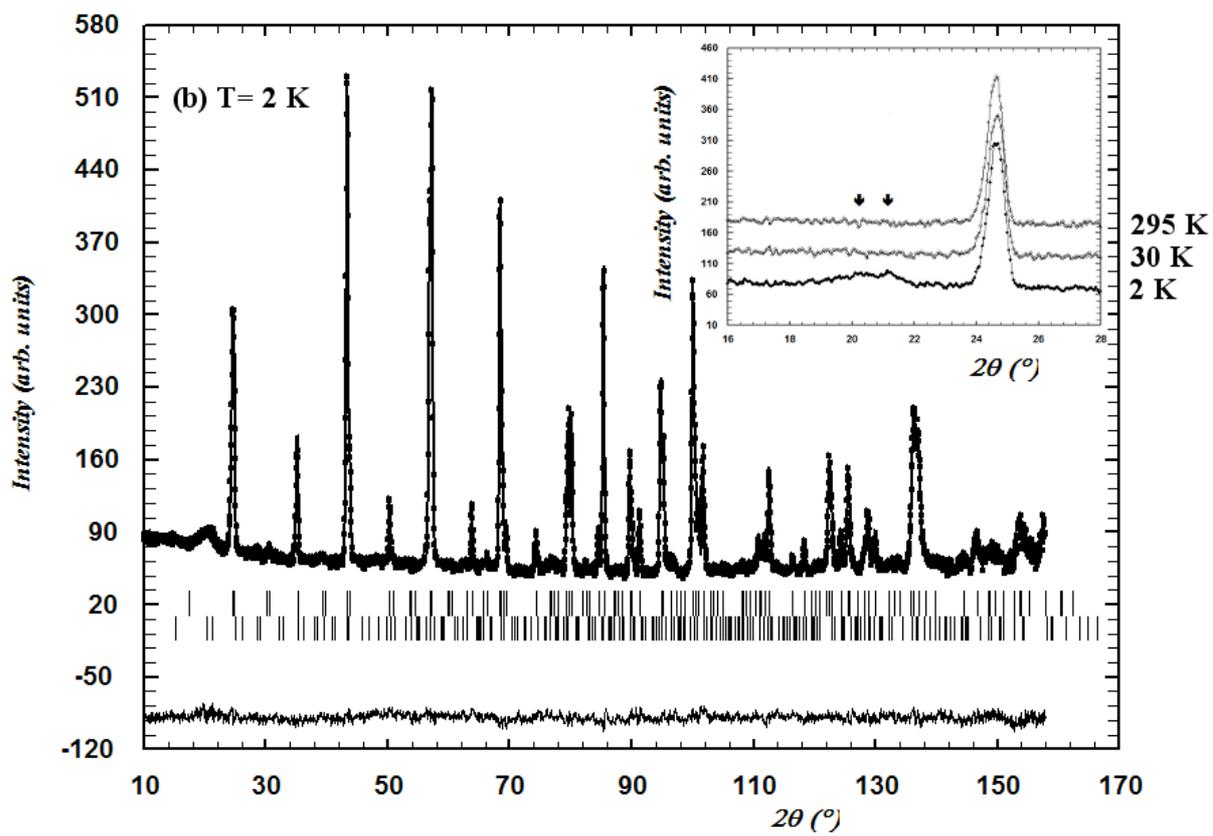

**Figure 5**



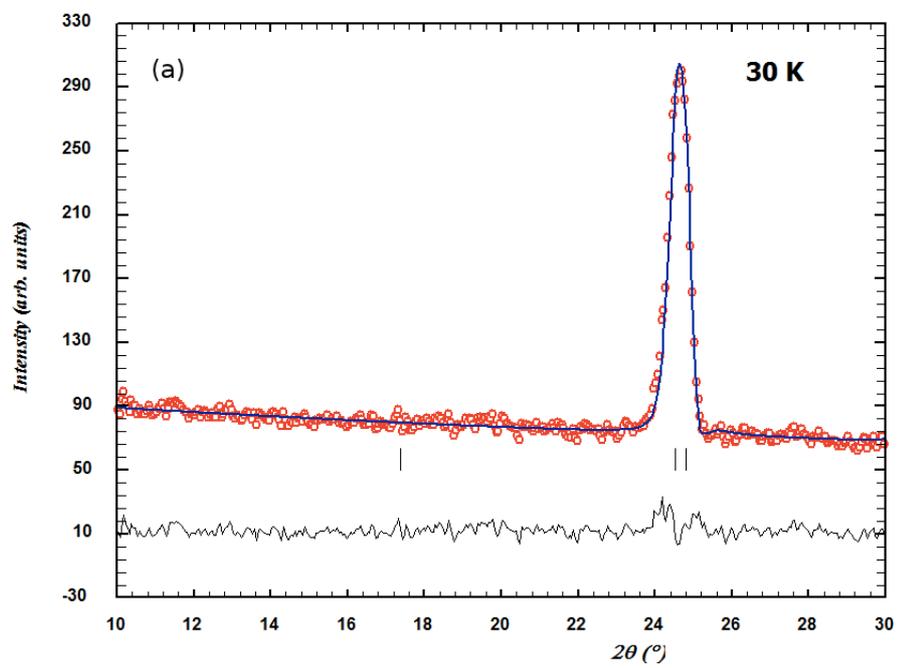

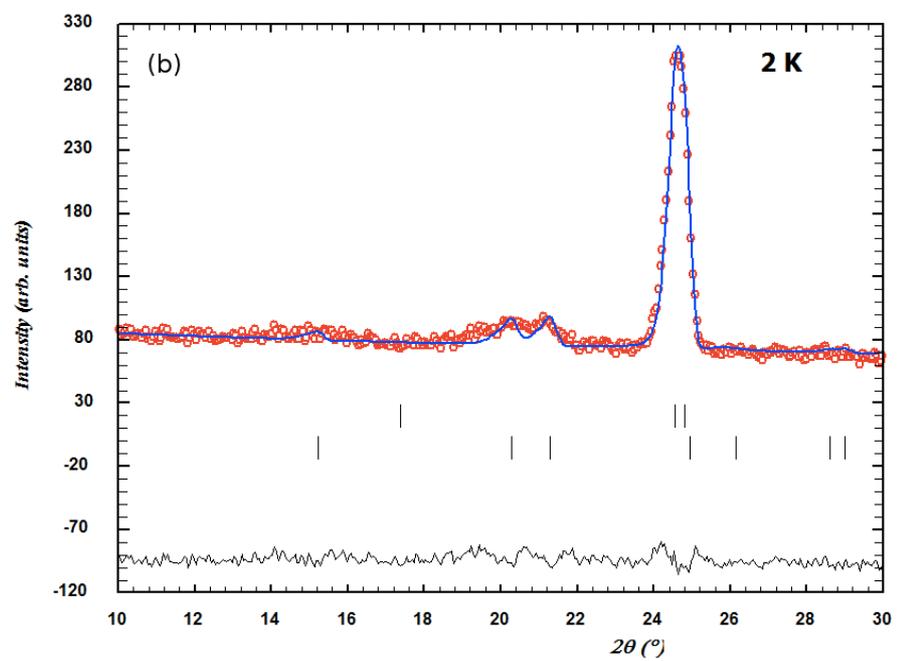

**Figure 6**



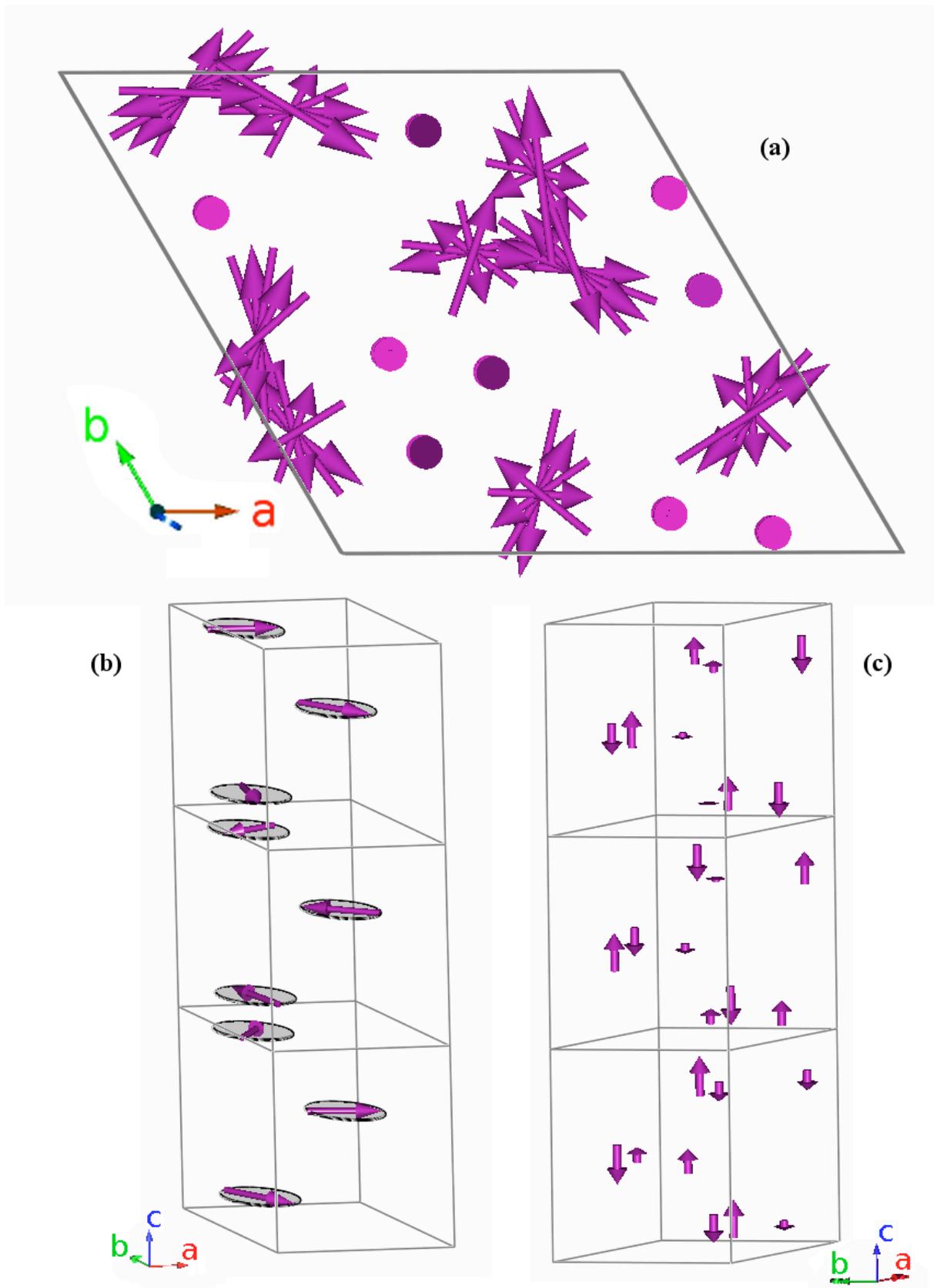

**Figure 7**



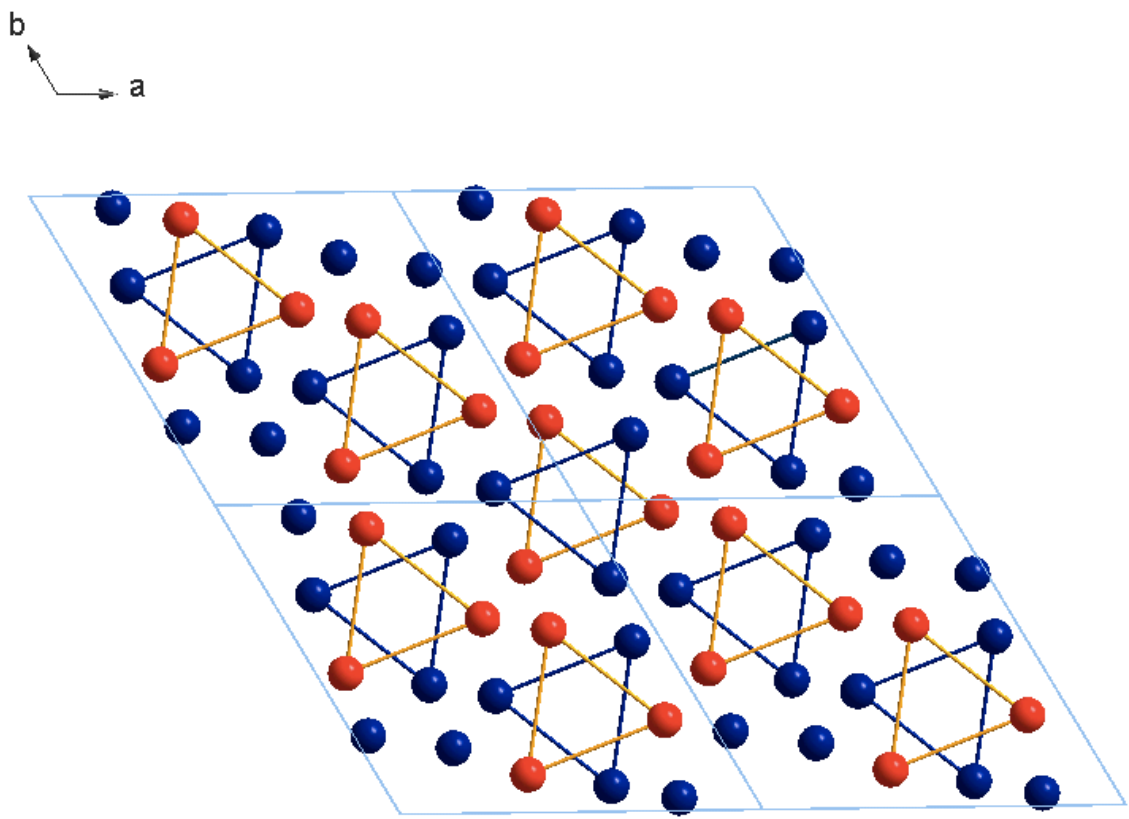

**Figure 8**



**Figure captions**

**Figure 1** The observed, calculated, and difference plots for the fit to the XRPD pattern of Mn$_{1.5}$$^{114}$Cd$_{1.5}$TeO$_6$ after the Rietveld refinement at 295K.

**Figure 2.** (a) Temperature dependence of the ZFC/FC susceptibility M/H of Mn$_{1.5}$$^{114}$Cd$_{1.5}$TeO$_6$, Mn$_2$CdTeO$_6$ and Mn$_3$TeO$_6$ recorded in low magnetic fields (50, 100 and 20 Oe respectively). (b) Temperature dependence of the ZFC/FC susceptibility of Mn$_{1.5}$$^{114}$Cd$_{1.5}$TeO$_6$ recorded in 5000 Oe. The inset shows the associated Curie-Weiss fit of the high-temperature data. The data obtained for MnCd$_2$TeO$_6$ is not included as the magnetization of this compound is dominated by additional spontaneous moments, reflecting tiny (not seen in XRD) amounts of Mn$_3$O$_4$ impurity phases.

**Figure 3.** Temperature dependence of the heat capacity C (plotted as C/T in the main frame and C in the inset) of Mn$_{1.5}$$^{114}$Cd$_{1.5}$TeO$_6$. The corresponding C/T data for Mn$_{1.5}$Cd$_{1.5}$TeO$_6$ is added for comparison.

**Figure 4** Polyhedral representation of the crystal structure of Mn$_{1.5}$$^{114}$Cd$_{1.5}$TeO$_6$.

**Figure 5** The observed, calculated and difference plots for the fit to the NPD patterns of Mn$_{1.5}$$^{114}$Cd$_{1.5}$TeO$_6$ after the Rietveld refinement of the nuclear and magnetic structure at 30 K (a), 2 K (b); the insert shows the temperature evolution of the magnetic reflections.

**Figure 6** Enlarged fragments of the NPD patterns at (a) 30 K and (b) 2 K and associated refinements. In (b), magnetic reflections are observed in addition to the nuclear ones. The lower row of peak positions mark those magnetic reflections.

**Figure 7** The magnetic structure of Mn$_{1.5}$$^{114}$Cd$_{1.5}$TeO$_6$: a) the elliptical envelope of the Mn(1) orbit perpendicular to the propagation vector; b) the variation of the magnetic moments in the Mn(1) sites along the propagation vector, c) the sine wave character of the Mn(2) orbit along the c-axis.

**Figure 8** Sketch of the crystal structure of Mn$_3$TeO$_6$ projected on the *ab* plane (using 2a x 2b x c lattice), in which only the Mn cations are represented and the sublattices of triangles are linked.